\documentclass[12pt]{article}
\usepackage[utf8]{inputenc}
\usepackage{lineno}
\usepackage{comment}
\usepackage{siunitx}
\usepackage{subcaption}
\usepackage{authblk} % for author/affiliation formatting
\usepackage[style=numeric,sorting=none]{biblatex}
\usepackage{graphicx}
\usepackage{amsmath}
\usepackage[english]{babel}

\title{Hybrid integration of two-dimensional dichalcogenides for low power saturable absorption in photonic integrated circuits}

\author[1,*]{Maria Carolina Volpato}
\author[1]{Gustavo H. Magro}
\author[1]{Antonio A.G. Von Zuben}
\author[1]{Luis A. M. Barea}
\author[2]{Pierre-Louis de Assis}
\author[1]{Newton C. Frateschi}

\affil[1]{\textit{"Gleb Wataghin" Institute of Physics, University of Campinas, Campinas, Brazil.}}
\affil[2]{\textit{Electrical Engineering Department, Universidade Federal de São Carlos (UFSCar), São Carlos, SP, Brazil.}}
\affil[*]{Corresponding author: m202733@dac.unicamp.br}

\date{} % remove date

\begin{document}

\maketitle
\begin{abstract} 

Silicon photonics provides a versatile platform for large-scale integration of optical functions, but its weak intrinsic nonlinear response limits the realization of active, intensity-dependent functionalities. Hybrid integration of two-dimensional (2D) materials has emerged as a promising strategy to overcome these limitations by enabling strong light–matter interaction and broadband absorption. Here, we demonstrate saturable absorption in a Complementary Metal – Oxide – Semiconductor (CMOS)-compatible silicon-on-insulator (SOI) microring resonator integrated with an exfoliated monolayer of 1T'-MoTe$_2$. Transmission measurements under varying input powers reveal a clear nonlinear absorption response, with a saturation power as low as \SI{2(1)}{\uW}. A phenomenological model accurately reproduces the experimental results, confirming the nonlinear behavior induced by the hybrid MoTe$_2$ integration. These findings establish a proof-of-concept for ultracompact, low-power saturable absorbers in photonic integrated circuits (PICs), paving the way for applications in integrated lasers, ultrafast optical signal processing, and neuromorphic photonics.

\end{abstract}

%%%%%%%%%%%%%%%%%%%%%%%%%%  body  %%%%%%%%%%%%%%%%%%%%%%%%%%
\section{Introduction}

Silicon Photonics has become a powerful platform for integrating optical functions at the chip scale, leveraging the compatibility of complementary metals-oxide-semiconductor (CMOS) and the technological maturity of microelectronics \cite{bogaerts2018silicon}. However, the intrinsic nonlinear properties of silicon are weak, which limits the implementation of active or intensity-dependent optical functionalities in silicon-based devices \cite{siew2021review,heebner2008optical}. This limitation is particularly critical for applications requiring efficient nonlinear optical responses, such as ultrafast switching \cite{bawankar2021microring} and pulse generation \cite{yu2021high}.

In this context, the hybrid integration of two-dimensional (2D) materials into photonic integrated circuits (PICs) has emerged as a promising strategy \cite{siew2021review,novoselov20162d}. These materials exhibit strong light–matter interaction, high single-layer absorption, and tunable electronic properties, offering a new degree of freedom in the design of optoelectronic devices \cite{peyskens2019integration,datta20242d}. Among the most relevant applications is saturable absorption (SA), a nonlinear effect in which optical absorption decreases as the incident light intensity increases \cite{bao2009atomic,yu2021high,reep2023active}. SA is essential for Q-switching, mode-locking in integrated lasers, and, more recently, photonic neural networks\cite{shen2017deep}.

Despite this strong potential, important challenges remain for the consolidation of 2D-material-based SAs in PICs: (i) environmental degradation-many crystalline phases, such as 1T'-MoTe$_2$, are unstable under ambient conditions and prone to oxidation, which reduces device functionality and reliability \cite{ruppert2014optical}; (ii) high saturation power-most reported works require input powers in the milliwatt range to observe the effect \cite{ma2019recent}; and (iii) compatibility with CMOS processes-essential for large-scale integration, but not always achieved in experimental demonstrations \cite{siew2021review}.

MoTe$_2$ in its 1T' phase stands out in this scenario due to its semimetallic character, which enables high modulation depth and theoretically low saturation intensity \cite{shen2017deep,volpato2023saturable}. However, experimental demonstrations of SAs in resonant cavities operating at Telecom wavelengths, especially on CMOS-compatible platforms, remain scarce, with limited discussion on exploiting the strong optical confinement of cavities to maximize light–matter interaction and enhance device performance. Regarding environmental degradation, SU-8 encapsulation after MoTe$_2$ transfer has not been widely reported for this material, but it is supported by studies showing SU-8 as an effective protective overlayer against oxidation in MoS$_2$-based devices\cite{kung2019air}, reinforcing the choice of this strategy.

In this work, we investigate the saturable absorption behavior of exfoliated 1T'-MoTe$_2$ monolayer heterogeneously integrated onto a silicon-on-insulator (SOI) microring resonator. The add-drop configuration enables precise evaluation of the transmission response as a function of input power. We experimentally demonstrate saturable absorption at ultra low power levels, with a saturation power of \SI{2(1)}{\micro W}, at least two orders of magnitude lower than in previous reports \cite{volpato2024analysis,volpato20251t}. This response was described using a phenomenological model, showing excellent agreement between theory and experiment.

\section{Experiment Methodology}

\subsection{Modelling and expected behavior} 
Ring resonators are widely employed in integrated photonics due to their compactness and versatility \cite{bawankar2021microring,datta20242d,souza2014embedded}. In an add-drop configuration, the field coupling can be described using a unitary operator acting on the wave amplitudes \cite{rabus2020ring,yariv2000universal}. For the symmetric add-drop configuration, we define an effective coupling coefficient $t$ such that $|t|^2 = r_1 r_2$, where $r_{1,2}$ are the self-coupling coefficients of the two couplers. The transmittance at the through port, $T_{thr}$, is then given by
\begin{equation}\label{eq:Transmitance A}
T_{\text{thr}} = \frac{P_{\text{out}}}{P_{\text{input}}} = \frac{A^2 |t|^2 + |t|^2 - 2A|t|^2\cos(\theta)}{1 + A^2 |t|^4 - 2A|t|^2\cos(\theta)},
\end{equation}
where $P_{\text{input}}$ and $P_{\text{out}}$ denote the input and output powers, respectively. Here, $t = |t|e^{i\phi_{t}}$ is the field coupling coefficient between the bus waveguide and the ring, and $A = e^{-\alpha L}$ is the attenuation factor, with $\alpha$ the propagation loss in $[1/m]$ and $L$ the cavity length. At resonance, $\theta \equiv n_{\text{eff}}2\pi \text{L}/\lambda = 2m\pi$, the through-port transmission reaches its minimum, while the stored power $P_{\text{str}}$ inside the resonator is maximum, given by
\begin{equation}\label{eq:Transmitance B}
\frac{P_{\text{str}}}{P_{\text{input}}} = \frac{A^2 |t|^2 (1 - |t|^2)}{(1-A|t|^2)^2},
\end{equation}
where $n_{\text{eff}}$ is the effective refractive index of the guided mode.

The presence of an absorption layer over the microring region will affect the attenuation factor. Therefore, if this layer is a saturable absorber, the transmittance spectrum, and the stored power, will depend on the input optical power. Using the phenomenological expression for the absorption, we can relate the absorption with the stored power as

\begin{equation}\label{eq:SA}  
\alpha = \frac{\alpha_{0}}{1+\frac{P_{\text{str}}}{P_{\text{sat}}}} + \alpha_{n},
\end{equation}
where $\alpha_{0}$ is the absorption coefficient for the portion of the ring with the 1T'-MoTe$_2$ overlayer for $P_{\text{str}}=0$, and $\alpha_{n}$ is the absorption coefficient of the remaining ring within the resonator. This expression will be used to estimate the saturation power of our device, $P_{\text{sat}}$, which is defined as the stored power that leads to half the absorption in the overlayer portion of the ring.

%The proposed  model is a simple phenomenological approximation of the non-linear effect added to the microring cavity. However, important insights are obtained by analyzing the system of equations \eqref{eq:Transmitance B} and \eqref{eq:SA}. This system can be solved by multiplying equation \eqref{eq:SA} by $-L = - 2\pi R$ where $R$ is the radius of the ring cavity and then take the exponential from both sides, then substitute $P_{str}$ from equation \eqref{eq:Transmitance B} and finally arriving at the implicit relation \eqref{eq:f}

By combining Eqs. \eqref{eq:SA} and \eqref{eq:Transmitance B}, we obtain a master equation for the system,

\begin{equation}\label{eq:f}
f(P_{\text{input}}, A) = A - A_{n}\exp\left(-\frac{\alpha_{0}L}{1+\frac{P_{\text{input}}(A^2 |t|^2 (1 - |t|^2))}{P_{\text{sat}} (1-A|t|^2)^2}} \right) = 0,
\end{equation}
where $A = A_{n}A_{0}$, with $A_{n} = e^{-\alpha_n L}$ representing the non-saturable loss and 
\begin{equation}
    A_{0} = \exp\left(-\frac{\alpha_{0}L}{1+P_{str}/P_{sat}}\right)
\end{equation} 
is the saturable component of the attenuation. This nonlinear equation can be solved numerically for $A$ and $P_{\text{input}}$, given a set of fixed parameters.

\begin{figure}[h!]
\centering
\includegraphics[trim={0 7cm 0cm 6cm},clip,width=\linewidth]{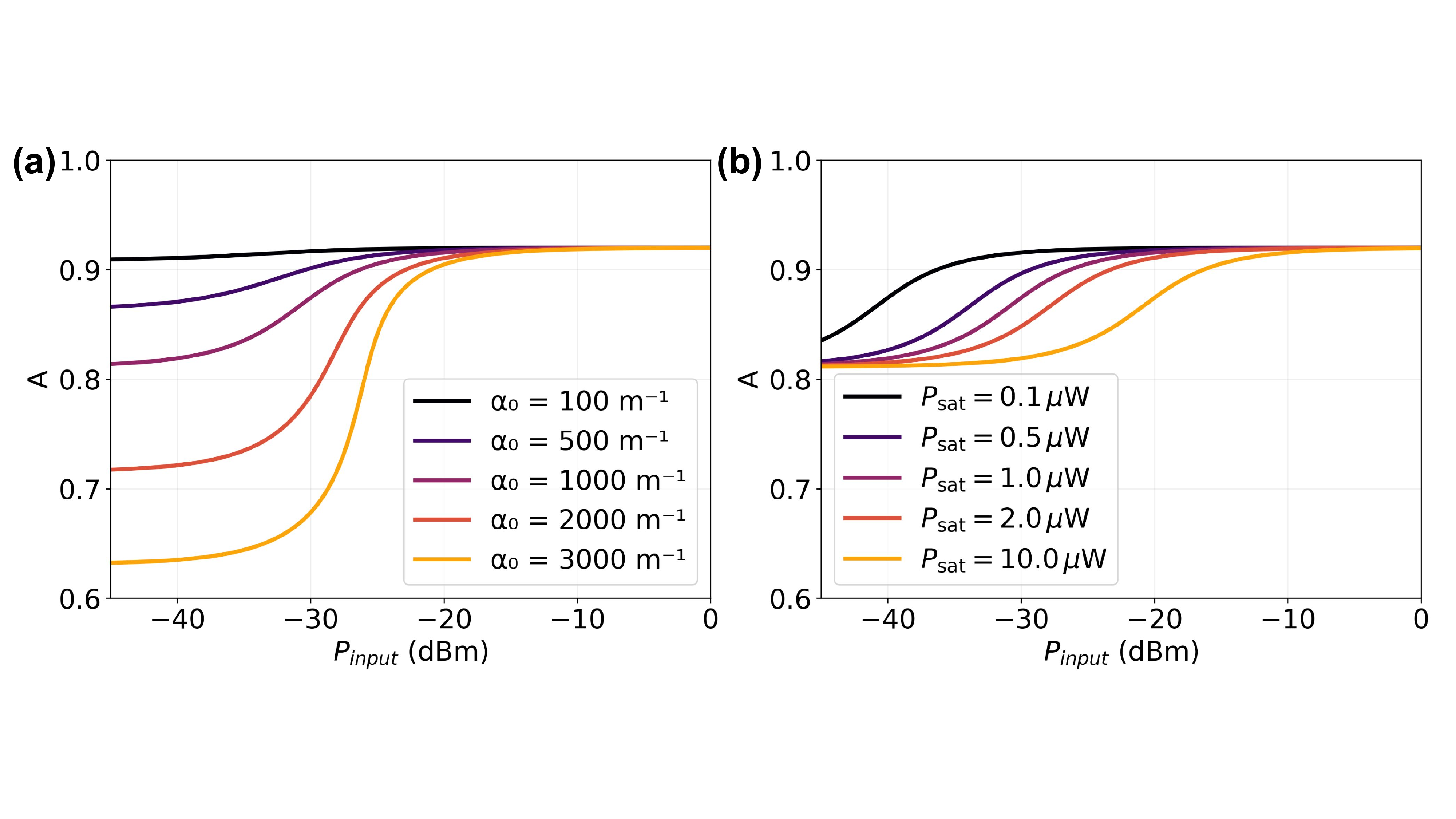}
\caption{Numerical solutions of Eq. \eqref{eq:f}: (a) varying $\alpha_{0}$ with constant $P_{\text{sat}}$ and constant $\alpha_n$, (b) varying $P_{\text{sat}}$ with constant $\alpha_{0}$ and constant $\alpha_n$.}
\label{fig:Sim_A}
\end{figure}

Figure \ref{fig:Sim_A}(a) shows the numerical solution of Eq. \eqref{eq:f} for a fixed saturation power $P_{\text{sat}} = \SI{1}{\micro\watt}$ and different values of the absorption coefficient $\alpha_0$; $\alpha_n$ is kept constant leading to a fixed $A_n=0.92$ . This value of $P_{\text{sat}}$ is consistent with the predictions in Ref. \cite{volpato2024analysis}. As $\alpha_0$ increases, the effective saturation power for P$_{\text{input}}$ increases slightly, shifting the asymptotic value of the attenuation factor. Notice that for a fixed $P_{\text{sat}}$, the stored power at saturation is always fixed. In contrast, Fig. \ref{fig:Sim_A}(b) presents the case where $\alpha_0 = \SI{1000}{m^{-1}}$ is fixed and $P_{\text{sat}}$ is varied; $\alpha_n$ is kept constant leading to a fixed $A_n=0.92$. As expected, increasing $P_{\text{sat}}$ requires proportionally higher input power to reach saturation.

%Regarding equation \eqref{eq:Transmitance A} when plotted in function of the wavelength,  no change in the curve will observed as we change the $P_{input}$ and a normalization is applied. However, as the material absorption $\alpha$ increases, the coefficient $A$ will decrease and the curve given by \eqref{eq:Transmitance A} will change its parameter $A$ that in turn will change its overall shape and then the system will become dependent on the input power. 

\begin{figure}[h!]
\centering
\includegraphics[trim={0 8cm 0cm 6cm},clip,width=\linewidth]{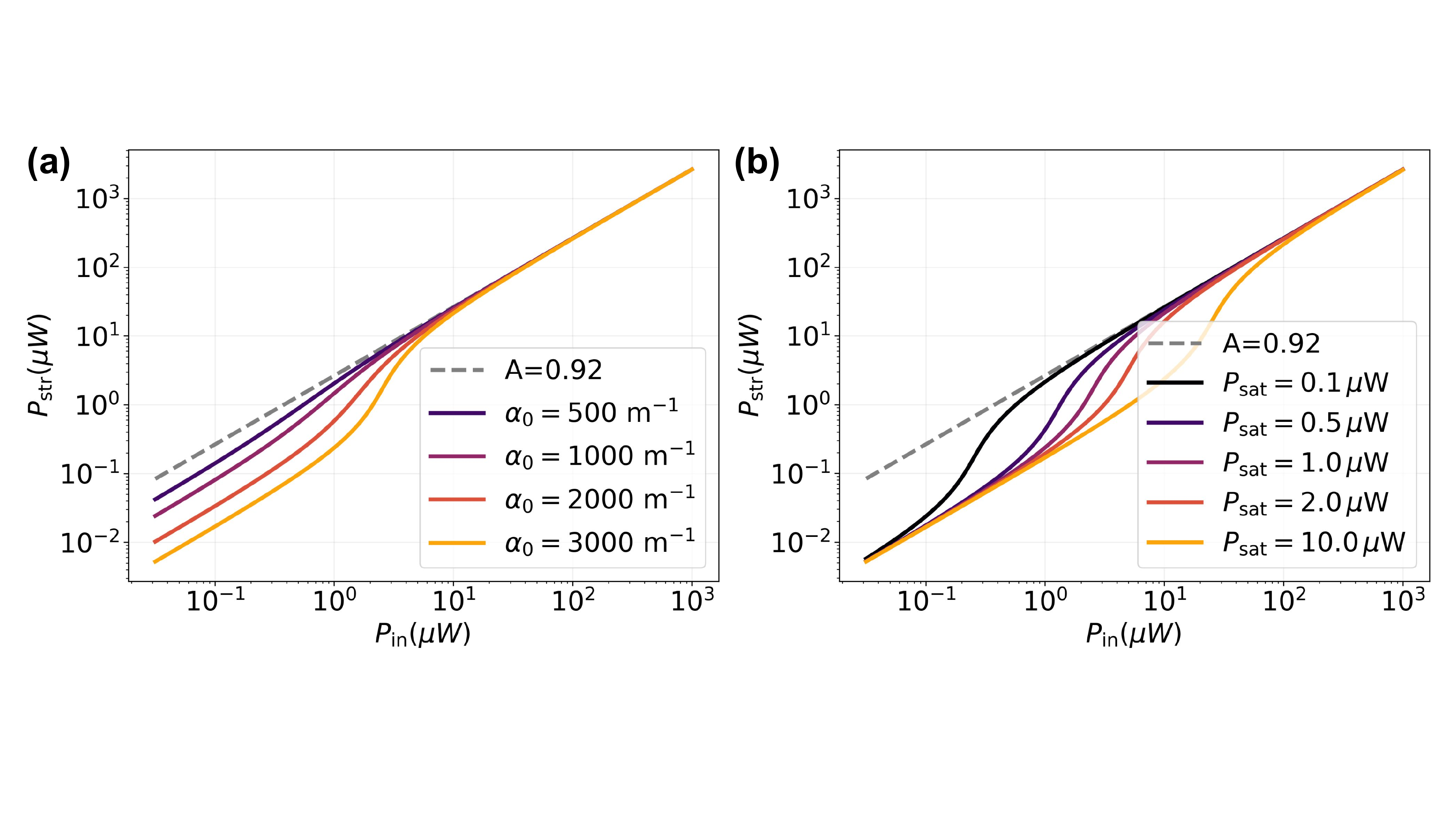}
\caption{Numerical solutions of Eq. \eqref{eq:Transmitance B} on the resonance: (a) varying $\alpha_{0}$ with constant $P_{sat}$, (b) varying $P_{sat}$ with constant $\alpha_{0}$.}
\label{fig:Sim_Pstr}
\end{figure}

Variations in input power also affect the stored power inside the cavity. As shown in Fig. \ref{fig:Sim_Pstr}, reducing the input power enhances the influence of absorption, leading to a rapid decrease in the stored energy. Figure \ref{fig:Sim_Pstr}(a) illustrates this effect for a fixed $P_{sat} = \SI{1}{\micro\watt}$ and different absorption coefficients $\alpha_0$, while Fig. \ref{fig:Sim_Pstr}(b) shows the case of fixed $\alpha_0$ and varying $P_{sat}$, which shifts the onset of the saturation regime to higher input powers.

A direct consequence of this behavior is a substantial change in the cavity quality factor. In this work, we define the total quality factor $Q_T$ as the ratio between the energy stored in the cavity and the loss per oscillation cycle \cite{ShaoqiSiliconPhotonics}. Since the quality factor is directly related to the photon lifetime, it decreases as the stored energy diminishes and the cavity losses increase-effects ultimately driven by the absorption introduced by the MoTe$_2$ layer.

The total quality factor can be separated into two contributions: the intrinsic quality factor $Q_0$, associated with intrinsic power losses (attenuation factor $A$), and the coupling quality factor $Q_c$, also denominated extrinsic quality factor, determined by the field coupling coefficient $t$. Their relation is given by
\begin{equation}\label{eq:Q}
\frac{1}{Q_T} = \frac{1}{Q_c} + \frac{1}{Q_0}.
\end{equation}

Expressing the transmittance in terms of the $Q$ parameters and the resonance frequency $\omega_0$ leads to the general form
\begin{equation}\label{eq:TransQ}
T(\omega) = \left| \frac{ \frac{1}{Q_0} - \frac{1}{Q_c} + \frac{i(\omega - \omega_0)}{\omega_0} }{\frac{1}{Q_0} + \frac{1}{Q_c} + \frac{i(\omega - \omega_0)}{\omega_0}} \right|^2.
\end{equation} This formulation allows $Q_0$ and $Q_c$ to be extracted directly from experimental data by fitting an isolated resonance.

\subsection{Sample Preparation}
\begin{figure}[h!]
\centerline{\includegraphics[width=0.6\textwidth]{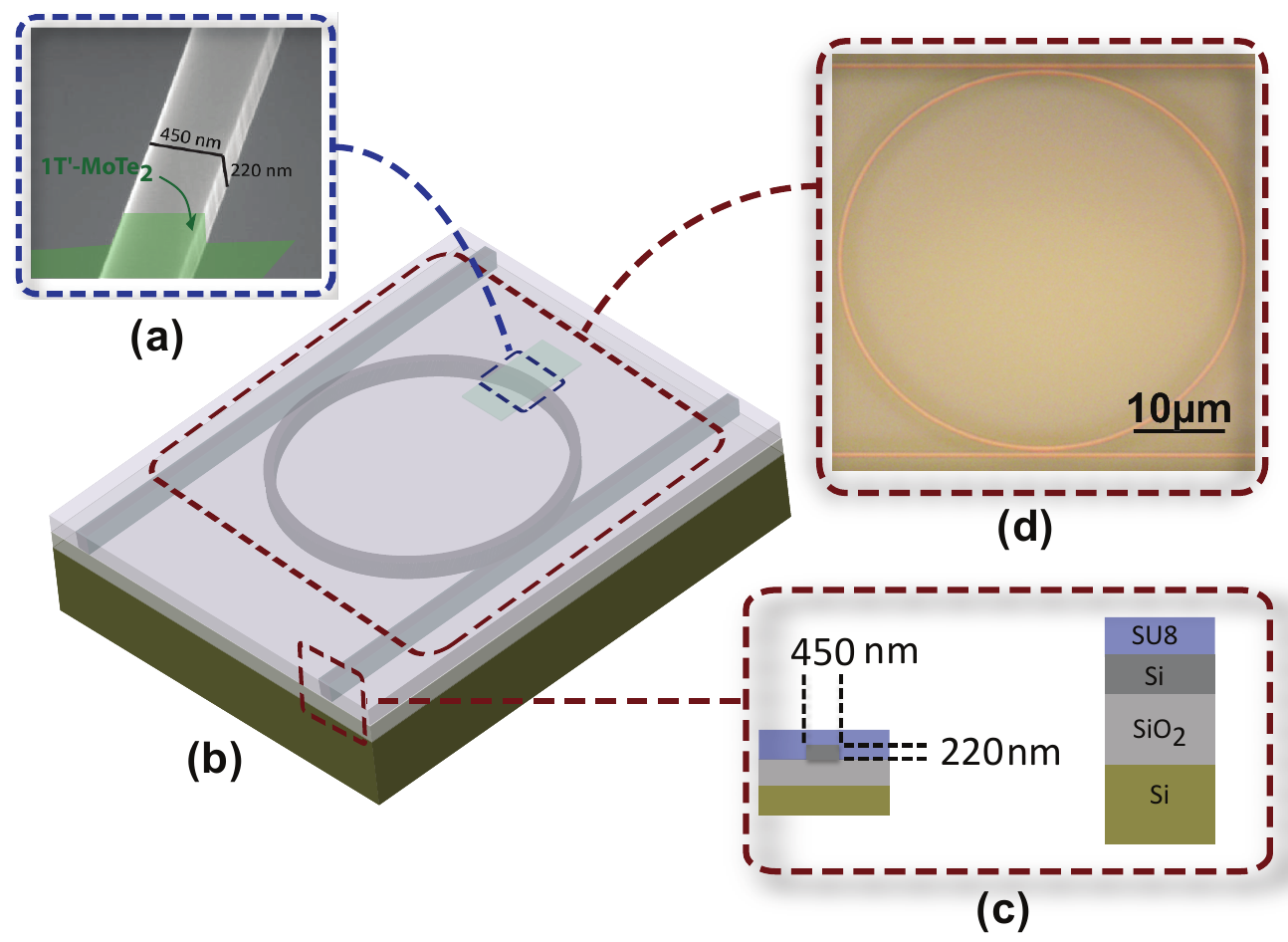}}
\caption{ (a) 3D schematic of the transferred MoTe$_2$ monolayer; (b) add-drop device; (c) cross-sectional schematic; (d) optical microscopy image prior to transfer.}
\label{fig:DescricaoSamplea}
\end{figure}

The photonic device employed in this study is presented in Fig. \ref{fig:DescricaoSamplea}. It was fabricated on SOI platform using standard CMOS-compatible processes at IMEC. The waveguide core consists of a \SI{450}{nm}-wide and \SI{220}{nm}-thick silicon layer, and the microring resonator has a radius of \SI{20}{\micro \meter}. The microring is coupled to both through and drop bus waveguides via symmetric coupling gaps of \SI{200}{nm}, designed to enable efficient power transfer. Fiber-to-chip coupling is facilitated by inverse nanotaper structures with tip widths of \SI{150}{nm}, enabling matching to lensed fibers. 

Transmission measurements were performed using a fiber-based coupling setup with a tunable IR laser set in the C-band \qtyrange[range-units=single,range-phrase=-]{1560}{1580}{\nm}, a polarization controller, and variable optical attenuator (VOA). A 90/10 beam splitter extracted a reference signal for monitoring, while a circulator minimized back-reflections. Lensed fibers with spot size $\approx$ \SI{2.8}{\um} are used to couple light into and out of the chip.

Initial characterization, hereafter named reference measurement, was conducted on SU-8 clad chips without MoTe$_2$, using an input power of \SI{8.98}{dBm} ($\approx$ \SI{7.9}{mW}). Power was progressively reduced in one step of \SI{20}{dB}, then steps of \SI{10}{dB} until the signal became undetectable.

Subsequently, the monolayer 1T'-MoTe$_2$ was mechanically exfoliated from bulk crystals using blue tape and transferred onto a polydimethylsiloxane (PDMS) substrate. Selected flakes were aligned and transferred onto the photonic strucures using a dry-transfer technique \cite{castellanos2014deterministic}. The transfer process was performed under ambient conditions with minimized air exposure to mitigate oxidation and degradation of the 1T' phase. After successful transfer, the devices were encapsulated with a SU-8 overlayer, applied via spin-coating and thermally cured. This encapsulation step serves both to protect the 2D material from environmental degradation and to provide mechanical robustness to the hybrid photonic device\cite{kung2019air}. 

\begin{figure}[h!]
    \centering
    \includegraphics[trim={0.5cm 8cm 1cm 6cm},clip,width=\linewidth]{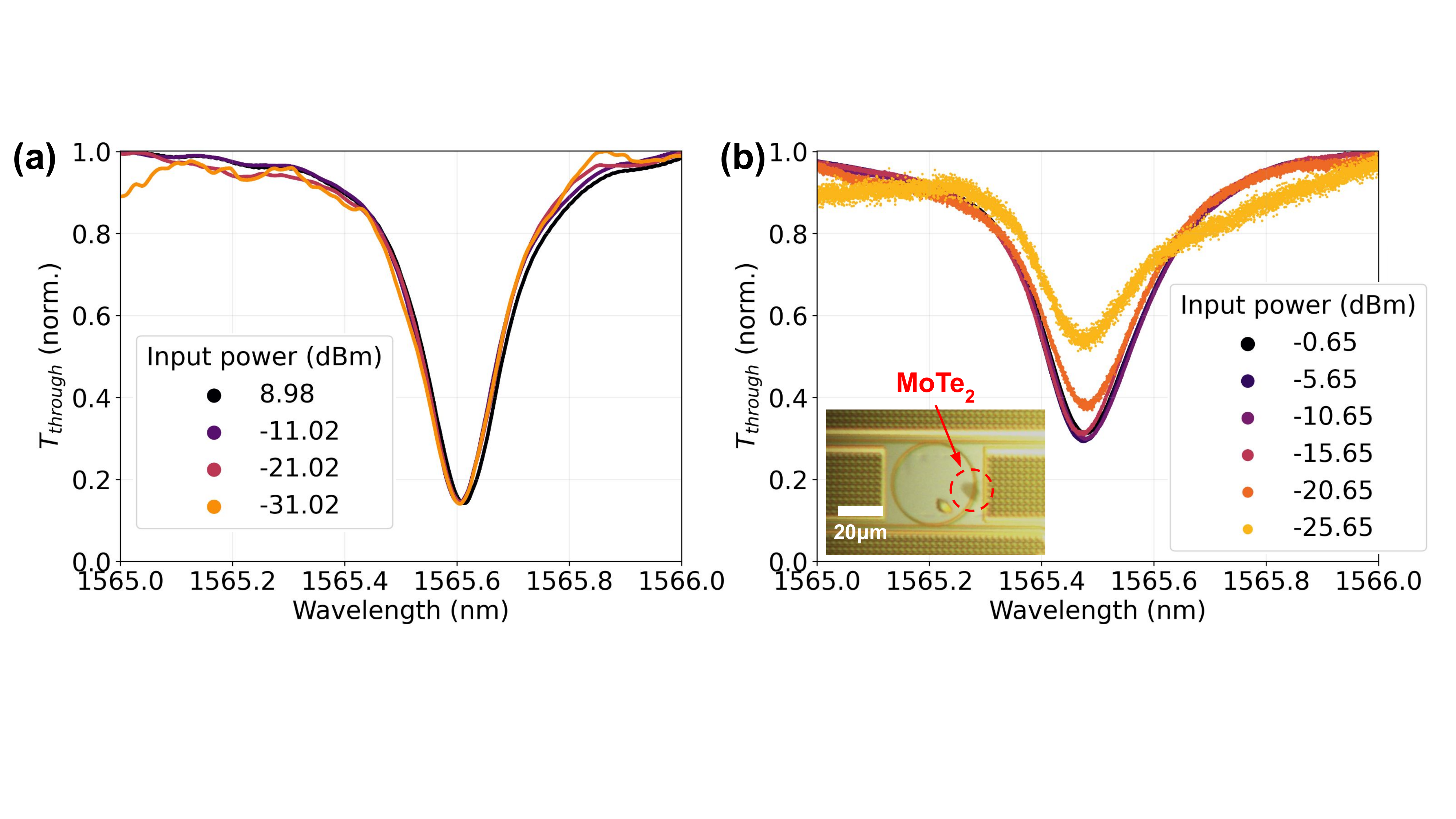}
    \caption{(a) Transmission spectra under different input power for the reference device (without 1T'-MoTe$_2$), used to extract $A$, $t$, and $n_{\text{eff}}$. (b) Transmission spectra for the device with the 1T'-MoTe$_2$ layer, used to estimate $P_{\text{sat}}$. Inset: Microscope image of the sample highlighting the 2D material.}
    \label{fig:measurements4}
\end{figure}

 The characterization of the MoTe$_2$ integrated devices followed a similar process as for the reference measurement. However, measurement started with a light input resulting in the lowest detectable transmission power. Subsequently, the light input power was increased in \SI{5}{dBm} steps to a maximum of \SI{-0.35}{dBm}. Input light polarization was fixed throughout each measurement.

\section{Discussion}

Fig. \ref{fig:measurements4}(a) shows the transmission measurement under different light input power for the device without 1T'-MoTe$_2$. As anticipated, variation in input power does not affect the transmission response. Fig. \ref{fig:measurements4}(b) shows the optical transmission experiment performed under different input light power for the sample with the 1T'-MoTe$_2$ layer. The latter graph reveals that the transmission spectrum remains unchanged until the input power drops below \SI{-15.65}{dBm}. At a power of \SI{-20.65}{dBm}, we observe a noticeable change in the spectrum.

\begin{figure}[h!]
    \centering
    \includegraphics[trim={0cm 9cm 0.5cm 6cm},clip,width=\linewidth]{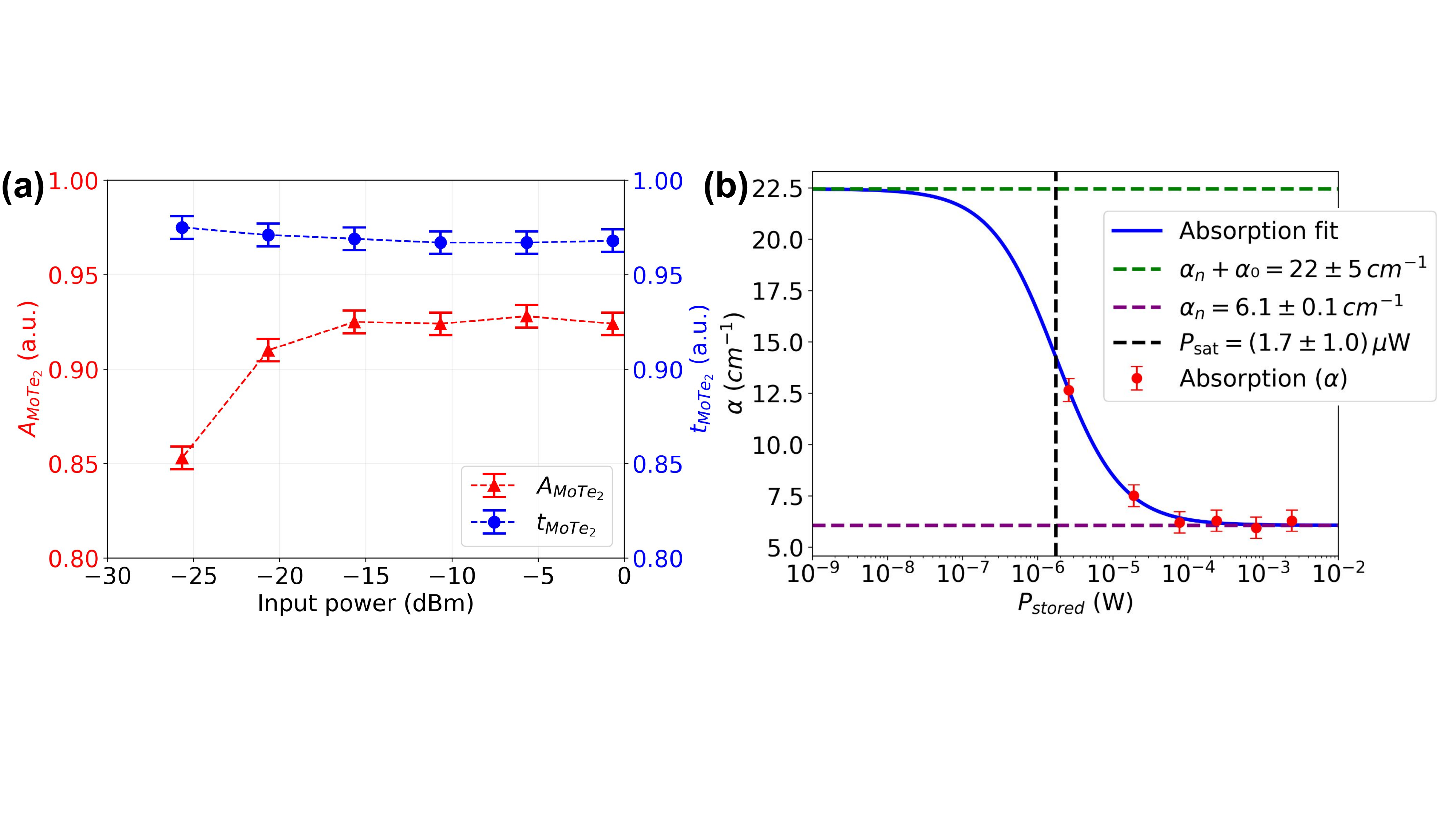}
    \caption{(a) Extracted attenuation $A$ and coupling $t$ vs. input power for the 1T'-MoTe$_2$-transferred device. (b) Fitted absorption curve using the experimental data.}
    \label{fig:measurements5}
\end{figure}

The observed transmission spectra dependence with input power shown in Fig. \ref{fig:measurements4}(b) can be explained by a dependence of the attenuation factor $A$ with power. In order to investigate this behavior, we used the data for both transmission experiment to fit the basic attenuation coefficient $A$ and the coupling factor $t$ using \eqref{eq:Transmitance A}; the fitting was performed for the data in a spectral window of $\pm$\SI{0.5}{nm} around the resonance wavelength. This window was chosen as the condition for convergence of the fitting extracted value. 

The reference measurements revealed minimal variation of the extracted parameters $A$ and $t$ with input power. However, Fig. \ref{fig:measurements5} (a) shows the fitting results for the transmission measurement with the transferred MoTe$_2$ layer where, clearly, the attenuation factor $A$ reduces as the input power drops below \SI{-15}{dB}, while the coupling coefficient $t$, remained essentially unchanged.  These results corroborate our hypothesis that the power dependence arises from attenuation variations due to saturable absorption since a reduction in attenuation factor $A$ is related to an increase in the absorption coefficient $\alpha$. 

To further investigate the variation in optical absorption with power, we used the experimental data to extrapolate the characteristic saturable absorption curve as described by Eq. \eqref{eq:SA}. Fig. \ref{fig:measurements5} (b) shows a fit of the saturable absorption curve from the experiment. We determined that $P_{\text{sat}}=$\SI{2(1)}{\uW}, which is consistent with our previous predictions \cite{volpato2024analysis}, and substantially lower than most reported experimental demonstrations \cite{volpato20251t}. 

%For a resonant cavity operating at steady-state without saturable absorber (SA) effects, the circulating power inside the cavity is significantly higher than the input power this is quantified by equation \eqref{eq:Transmitance B}. 

The presence of saturable absorption modifies the cavity response, leading to a power-dependent quality factor. To investigate this dependence, a dedicated analysis was conducted. This was done by fitting the transmittance data to equation \eqref{eq:TransQ} using a $\pm$\SI{0.5}{nm} wavelength window. 

\begin{figure}[h!]
    \centering
    \includegraphics[width=0.8\linewidth]{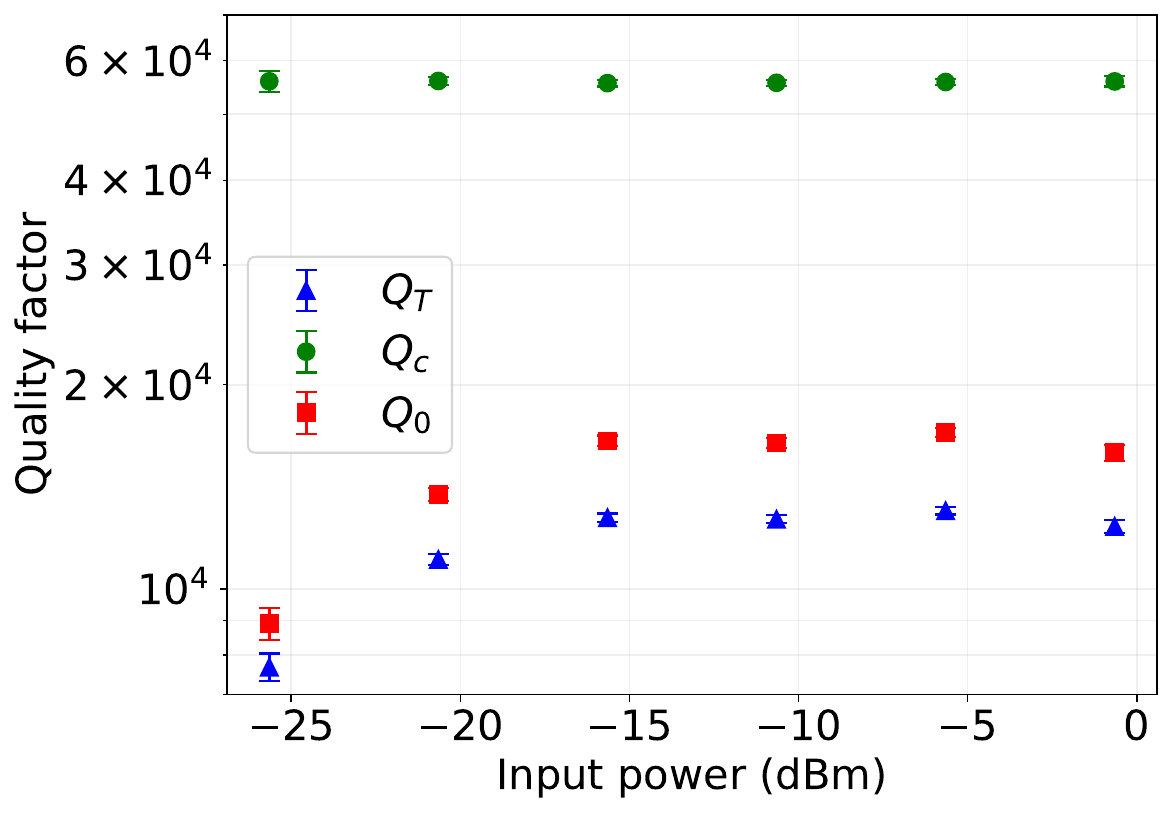}
    \caption{Quality factors $Q$, $Q_0$, $Q_c$ extracted from fitting equation \eqref{eq:TransQ} in the transmission data from the SA transferred device.}
    \label{fig:AQ}
\end{figure}

Fig. \ref{fig:AQ} shows the extracted values where we separated the contributions from the internal (intrinsic) and the external (extrinsic) quality factors.The extrinsic quality factor, $Q_e$ (related to the coupling coefficient $t$), remains constant as expected. In contrast, the intrinsic quality factor, $Q_0$ (related to absorption), changes significantly, decreasing as the input power is lowered. Consequently, the total quality factor of the cavity, $Q_{tot}$, also decreases. This demonstrates the connection between the stored energy model (given by Eq. \eqref{eq:Transmitance B} and illustrated in Fig. \ref{fig:Sim_Pstr}) and the cavity's quality factors. As the input power decreases, saturable absorption occurs, causing a rapid reduction in stored energy. This drop in energy directly implies a reduction in the cavity's quality factor. In particular, for the condition where $\alpha/2=\alpha_n+\alpha_0 =$ \SI{14.26}{cm^{-1}} (i.e., at $P_{\text{str}}=P_{\text{sat}}$), the stored power reaches about \SI{91}{\percent} of the input power $P_{\text{input}}$. This also explains why the observed spectral modifications appear around input levels of $-15$ to \SI{-20}{dBm}, even though the extracted $P_{\text{sat}}$ is on the order of only a few microwatts: the relevant parameter is the stored intracavity power $P_{\text{str}}$, which can greatly exceed the launched input power.

We noticed that the result should be different if the layer was placed over the coupling region.  On the other hand, the intrinsic quality factor, $Q_0$ which is related to the ring losses, undergoes a substantial increase as the input power is reduced.  Also, as expected from equation \eqref{eq:Q}, since $Q_0 << Q_c$ we obtain that $Q \approx Q_0$ and $Q$ follows a similar behavior as $Q_0$.

\section{Conclusion}
We demonstrated the hybrid integration of a monolayer 1T'-MoTe$_2$ saturable absorber onto a SOI microring resonator in an add-drop configuration. A phenomenological model accurately captured the power-dependent transmission, enabling the extraction of key absorption parameters. The estimated saturation power of \SI{2(1)}{\micro \watt} represents a substantial reduction compared to most previously reported devices. These findings establish a proof-of-concept for the integration of two-dimensional materials into photonic circuits and emphasize their potential for ultra-compact, CMOS-compatible, and energy-efficient nonlinear optical devices.

\section*{Funding}
Conselho Nacional de Desenvolvimento Científico e  Tecnológico (440231/2021-3,305282/2022-0,302959/2025-4); Fundação de Amparo à Pesquisa do Estado de São Paulo (2018/25339-4,2024/08855-0), Air Force Office of Scientific Research (AFOSR) (FA9550-20-1-0002).
\section*{Acknowledgment}
The authors would like to thank Dr. Ingrid D. Barcelos, Dr. Alisson R. Cadore, and Prof. Henrique G. Rosa for the fruitful discussions during the preparation of this manuscript.

\section*{Disclosures}
The authors declare no conflicts of interest.

\section*{Data availability} 
Data underlying the results presented in this paper are not publicly available at this time but may be obtained from the authors upon reasonable request.
\printbibliography

@article{siew2021review,
  title={Review of silicon photonics technology and platform development},
  author={Siew, Shawn Yohanes and Li, Bo and Gao, Feng and Zheng, Hai Yang and Zhang, Wenle and Guo, Pengfei and Xie, Shawn Wu and Song, Apu and Dong, Bin and Luo, Lian Wee and others},
  journal={Journal of Lightwave Technology},
  volume={39},
  number={13},
  pages={4374--4389},
  year={2021},
  publisher={OSA}
}

@book{heebner2008optical,
  title={Optical microresonators: theory, fabrication, and applications},
  author={Heebner, John and Grover, Rohit and Ibrahim, Tarek},
  year={2008},
  publisher={Springer}
}

@inproceedings{bawankar2021microring,
  title={Microring resonators based applications in silicon photonics-a review},
  author={Bawankar, Yash R and Singh, Anamika},
  booktitle={2021 5th conference on information and communication technology (CICT)},
  pages={1--6},
  year={2021},
  organization={IEEE}
}

@article{yu2021high,
  title={High-yield exfoliation of monolayer 1T’-MoTe2 as saturable absorber for ultrafast photonics},
  author={Yu, Wei and Dong, Zikai and Abdelwahab, Ibrahim and Zhao, Xiaoxu and Shi, Jia and Shao, Yan and Li, Jing and Hu, Xiao and Li, Runlai and Ma, Teng and others},
  journal={ACS nano},
  volume={15},
  number={11},
  pages={18448--18457},
  year={2021},
  publisher={ACS Publications}
}

@inproceedings{volpato2024analysis,
  title={Analysis 1T'-MoTe 2 Saturable Absorber Integrated to a Silicon Nitride Waveguide},
  author={Volpato, Maria Carolina and Rosa, Henrique G and de Assis, Pierre-Louis and Frateschi, Newton Ces{\'a}rio},
  booktitle={2024 SBFoton International Optics and Photonics Conference (SBFoton IOPC)},
  pages={1--3},
  year={2024},
  organization={IEEE}
}

@article{shen2017deep,
  title={Deep learning with coherent nanophotonic circuits},
  author={Shen, Yichen and Harris, Nicholas C and Skirlo, Scott and Prabhu, Mihika and Baehr-Jones, Tom and Hochberg, Michael and Sun, Xin and Zhao, Shijie and Larochelle, Hugo and Englund, Dirk and others},
  journal={Nature photonics},
  volume={11},
  number={7},
  pages={441--446},
  year={2017},
  publisher={Nature Publishing Group UK London}
}

@article{kung2019air,
  title={Air and water-stable n-type doping and encapsulation of flexible MoS2 devices with SU8},
  author={Kung, Yen-Cheng and Hosseini, Nahid and Dumcenco, Dumitru and Fantner, Georg E and Kis, Andras},
  journal={Advanced Electronic Materials},
  volume={5},
  number={1},
  pages={1800492},
  year={2019},
  publisher={Wiley Online Library}
}

@article{volpato20251t,
  title={1T'-MoTe $ \_2 $ as an integrated saturable absorber for photonic machine learning},
  author={Volpato, Maria Carolina and Rosa, Henrique G and Reep, Tom and de Assis, Pierre-Louis and Frateschi, Newton Cesario},
  journal={arXiv preprint arXiv:2507.16140},
  year={2025}
}

@inproceedings{volpato2023saturable,
  title={Saturable absorption in the C-Band employing 2D IT’-MoTe 2},
  author={Volpato, Maria Carolina and Rosa, HG and De Assis, PL and Frateschi, N Ces{\'a}rio},
  booktitle={2023 IEEE Photonics Conference (IPC)},
  pages={1--2},
  year={2023}
}

@article{bogaerts2018silicon,
  title={Silicon photonics circuit design: methods, tools and challenges},
  author={Bogaerts, Wim and Chrostowski, Lukas},
  journal={Laser \& Photonics Reviews},
  volume={12},
  number={4},
  pages={1700237},
  year={2018},
  publisher={Wiley Online Library}
}

@article{novoselov20162d,
  title={2D materials and van der Waals heterostructures},
  author={Novoselov, K S and Mishchenko, Artem and Carvalho, Alexandra and Castro Neto, AH},
  journal={Science},
  volume={353},
  number={6298},
  pages={aac9439},
  year={2016},
  publisher={American Association for the Advancement of Science}
}

@article{peyskens2019integration,
  title={Integration of single photon emitters in 2D layered materials with a silicon nitride photonic chip},
  author={Peyskens, Fr{\'e}d{\'e}ric and Chakraborty, Chitraleema and Muneeb, Muhammad and Van Thourhout, Dries and Englund, Dirk},
  journal={Nature communications},
  volume={10},
  number={1},
  pages={4435},
  year={2019},
  publisher={Nature Publishing Group UK London}
}

@article{datta20242d,
  title={2D material platform for overcoming the amplitude--phase tradeoff in ring resonators},
  author={Datta, Ipshita and Gil-Molina, Andres and Chae, Sang Hoon and Zhou, Vivian and Hone, James and Lipson, Michal},
  journal={Optica},
  volume={11},
  number={1},
  pages={48--57},
  year={2024},
  publisher={Optica Publishing Group}
}

@inproceedings{reep2023active,
  title={Active and passive mode-locking of a laser using a graphene modulator on an SOI chip},
  author={Reep, T and Wu, C and Brems, S and Yudistira, D and Van Campenhout, J and Pantouvaki, M and Van Thourhout, D and Kuyken, B},
  booktitle={2023 IEEE Photonics Conference (IPC)},
  pages={1--2},
  year={2023},
  organization={IEEE}
}

@article{ruppert2014optical,
  title={Optical properties and band gap of single-and few-layer MoTe2 crystals},
  author={Ruppert, Claudia and Aslan, Burak and Heinz, Tony F},
  journal={Nano letters},
  volume={14},
  number={11},
  pages={6231--6236},
  year={2014},
  publisher={ACS Publications}
}

@article{bao2009atomic,
  title={Atomic-layer graphene as a saturable absorber for ultrafast pulsed lasers},
  author={Bao, Qiaoliang and Zhang, Han and Wang, Yu and Ni, Zhenhua and Yan, Yongli and Shen, Ze Xiang and Loh, Kian Ping and Tang, Ding Yuan},
  journal={Advanced Functional Materials},
  volume={19},
  number={19},
  pages={3077--3083},
  year={2009},
  publisher={Wiley Online Library}
}

@article{ma2019recent,
  title={Recent progress in ultrafast lasers based on 2D materials as a saturable absorber},
  author={Ma, Chunyang and Wang, Cong and Gao, Bo and Adams, Jordan and Wu, Ge and Zhang, Han},
  journal={Applied Physics Reviews},
  volume={6},
  number={4},
  year={2019},
  publisher={AIP Publishing}
}

@incollection{rabus2020ring,
  title={Ring resonators: Theory and modeling},
  author={Rabus, Dominik Gerhard and Sada, Cinzia},
  booktitle={Integrated Ring Resonators: A Compendium},
  pages={3--46},
  year={2020},
  publisher={Springer}
}

@article{yariv2000universal,
  title={Universal relations for coupling of optical power between microresonators and dielectric waveguides},
  author={Yariv, Amnon},
  journal={Electronics letters},
  volume={36},
  number={4},
  pages={321--322},
  year={2000},
  publisher={IET}
}

@article{ShaoqiSiliconPhotonics,
  title={Silicon photonics: from a microresonator perspective},
  author={Shaoqi and Feng, Ting and Lei, Hui and Chen, Hong and Cai, Xianshu and Luo and Andrew W. Poon },
  journal={Laser Photonics},
  volume={6},
  number={2},
  pages={ 145--177},
  year={2012},
  publisher={Wiley}
}

@article{castellanos2014deterministic,
  title={Deterministic transfer of two-dimensional materials by all-dry viscoelastic stamping},
  author={Castellanos-Gomez, Andres and Buscema, Michele and Molenaar, Rianda and Singh, Vibhor and Janssen, Laurens and Van Der Zant, Herre SJ and Steele, Gary A},
  journal={2D Materials},
  volume={1},
  number={1},
  pages={011002},
  year={2014},
  publisher={IOP Publishing}
}

@article{souza2014embedded,
  title={Embedded coupled microrings with high-finesse and close-spaced resonances for optical signal processing},
  author={Souza, Mario CMM and Barea, Luis AM and Vallini, Felipe and Rezende, Guilherme FM and Wiederhecker, Gustavo S and Frateschi, Newton C},
  journal={Optics express},
  volume={22},
  number={9},
  pages={10430--10438},
  year={2014},
  publisher={Optical Society of America}
}
\end{document}